\documentclass[conference]{IEEEtran}
\IEEEoverridecommandlockouts
\usepackage{cite}
\usepackage{amsmath,amssymb,amsfonts}
\usepackage{algorithmic}
\usepackage{graphicx}
\usepackage{float}
\usepackage{subfigure}
\usepackage{textcomp}
\usepackage{xcolor}
\def\BibTeX{{\rm B\kern-.05em{\sc i\kern-.025em b}\kern-.08em
    T\kern-.1667em\lower.7ex\hbox{E}\kern-.125emX}}
\begin{document}

\title{Light Field Compression Based on Implicit Neural Representation}

\author{\IEEEauthorblockN{Henan Wang, Hanxin Zhu, Zhibo Chen*\thanks{Zhibo Chen is the corresponding author. (chenzhibo@ustc.edu.cn)}}
\IEEEauthorblockA{\textit{CAS Key Laboratory of Technology in Geo-spatial Information Processing and Application System} \\
\textit{University of Science and Technology of China}\\
Hefei, China \\
chenzhibo@ustc.edu.cn}
}

\maketitle

\begin{abstract}
Light field, as a new data representation format in multimedia, has the ability to capture both intensity and direction of light rays. However, the additional angular information also brings a large volume of data. Classical coding methods are not effective to describe the relationship between different views, leading to redundancy left. To address this problem, we propose a novel light field compression scheme based on implicit neural representation to reduce redundancies between views. We store the information of a light field image implicitly in an neural network and adopt model compression methods to further compress the implicit representation. Extensive experiments have demonstrated the effectiveness of our proposed method, which achieves comparable rate-distortion performance as well as superior perceptual quality over traditional methods.
\end{abstract}


\section{Introduction}
Light field is a new data representation in multimedia that can record both the intensity and direction of light rays. The 2-D array format of the light field can be regarded as pictures captured from different viewpoints. Each view, namely sub-aperture image (SAI), is indexed by the angular coordinate $(u,v)$. With each SAI containing two spatial dimensions $(x,y)$, the light field contains four-dimensional (4-D) information. Due to the extra dimensions compared to traditional 2-D images, light field images are able to create more immersive experiences and extra degrees of freedom (DoFs) in different scenarios, such as virtual reality (VR), augmented reality (AR) and 3D displays\cite{conti2020dense}. However, the light field also carries a sheer amount of data, bringing huge challenges for its transmission or storage. Therefore, a more efficient representation for light fields is in high demand.

In recent years, implicit neural representation (INR) has gained increasing attention because it is a continuous and compact representation with many advantages over traditional discrete representations such as meshes, voxels and point clouds. As a rule, INR is realized by training a neural network to learn a mapping from input coordinates to scene properties such as pixel colors \cite{chen2021nerv}, volume densities \cite{mildenhall2020nerf} and signed distances \cite{park2019deepsdf}. With such a representation, many downstream tasks like denoising \cite{pearl2022nan}, super-resolution \cite{chen2022videoinr} and compression \cite{dupont2021coin} can be achieved, promising its superiority. In this paper, we propose to apply INR to light fields for compression. 

\begin{figure}[t]
    \centering
	\includegraphics[width=0.45\textwidth]{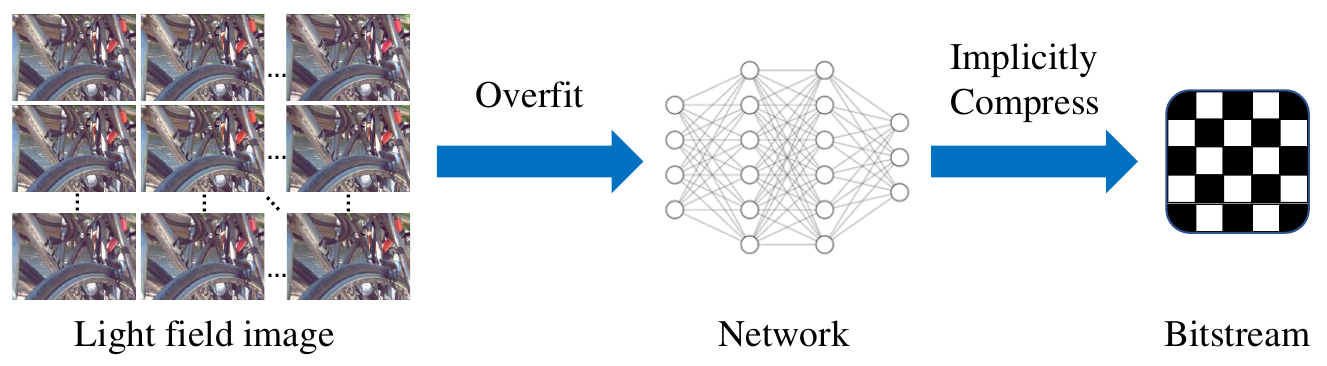}
    \caption{The encoding pipeline of our method. The INR network is well optimized to overfit the target light field image, then implicitly compressed to the bitstream.}
    \label{fig:pipeline}
\end{figure}

Different from the typical end-to-end compression frameworks that acquire and compress the latent representation\cite{balle2016end}, we compress light field images in a novel way by virtue of implicit neural representation. The information of a light field is stored in parameters of the INR network by overfitting. The parameters can be compressed using existing model compression techniques. After model compression, the bitstream containing compressed network parameters is acquired. The decoder recovers the INR network and then reconstructs the image. Since the encoding procedure is equivalent to network training, our method does not require datasets or a long training time for generalization. The whole pipeline is shown in Figure \ref{fig:pipeline}.

In this paper, we design an SAI-wise INR model which is able to generate a sub-aperture image every inference. We compare the rate-distortion performance and perceptual quality of our proposed method with other mainstream light field compression solutions. Experiments demonstrate that our method can provide superior visual quality and better performance at high bitrates. In addition, with the use of implicit representation, our model has greater decoding flexibility. This enables us to reconstruct an arbitrary viewpoint rather than recovering the entire light field image, which results in improved region of interest (ROI) functionalities.

Our contributions are described in the following:

\begin{itemize}
    \item  We propose a novel implicit representation model for light fields. Compared to pixel-wise INR models, our SAI-wise model is more efficient in training and inference.
    \item  We apply our light field INR model for the compression task. Experiments show that our compression performance is competitive against existing light field and INR-based compression methods.
\end{itemize}

\section{Related Work}
\label{sec:related work}
\subsection{Light Field Compression}
Due to the massive size of the light field, various light field compression strategies have been developed so far. The light field image may be interpreted as either a typical 2-D image (lenslet format)\cite{levoy1996light}, or pseudo video sequence (PVS) by arranging SAIs in a certain order\cite{vieira2015data} and compressed by image codecs like JPEG or video codecs like HEVC and VVC. However, because these algorithms are unable to effectively leverage the 4-D correlations of light field images, the compression performance is poor. Other compression approaches, such as 4-D transform based coding\cite{de20184d}, non-local spatial prediction based coding\cite{li2015coding}, multi-view based coding\cite{ahmad2017interpreting} and view synthesis based coding\cite{astola2018wasp,zhao2017light} have been proposed to fully exploit correlations among all dimensions of the light field. Recently, deep learning-based compression methods have achieved great success. Various end-to-end light field compression pipelines\cite{tong2022sadn} have also been proposed.

\subsection{Implicit Neural Representation}
Thanks to the success of NeRF, implicit neural representation (INR) has demonstrated its great power in many fields. \cite{mildenhall2020nerf} proposed to combine INR and neural rendering for photo-realistic novel view synthesis and 3D scene reconstruction. \cite{chen2022videoinr} took advantage of the continuous characteristic of INR for better video super-resolution. Image and video compression can also be realized by INR with some other model compression technologies. Considering the strong representation ability of CLIP, \cite{wang2022clip} proposed CLIP-NeRF, a multi-modal 3D object manipulation method for neural radiance fields. Moreover, INR can also be used for some generation tasks with the help of models such as GAN (Generative adversarial networks) and VAE (Variational autoencoder). Different from most INR-based methods that construct a pixel-wise mapping from pixel coordinates to pixel colors, NeRV is proposed to utilize an image-wise INR for faster training and inference process.

\section{Proposed Method}
We first introduce the implicit representation for light fields in Section \ref{sec:INR-LF}, including positional encoding, network architecture, loss objective and so on. Then we present the model compression method for the INR network in Section \ref{sec:modelcomp}.

\begin{figure*}[ht]
    \centering
    \includegraphics[width=0.9\textwidth]{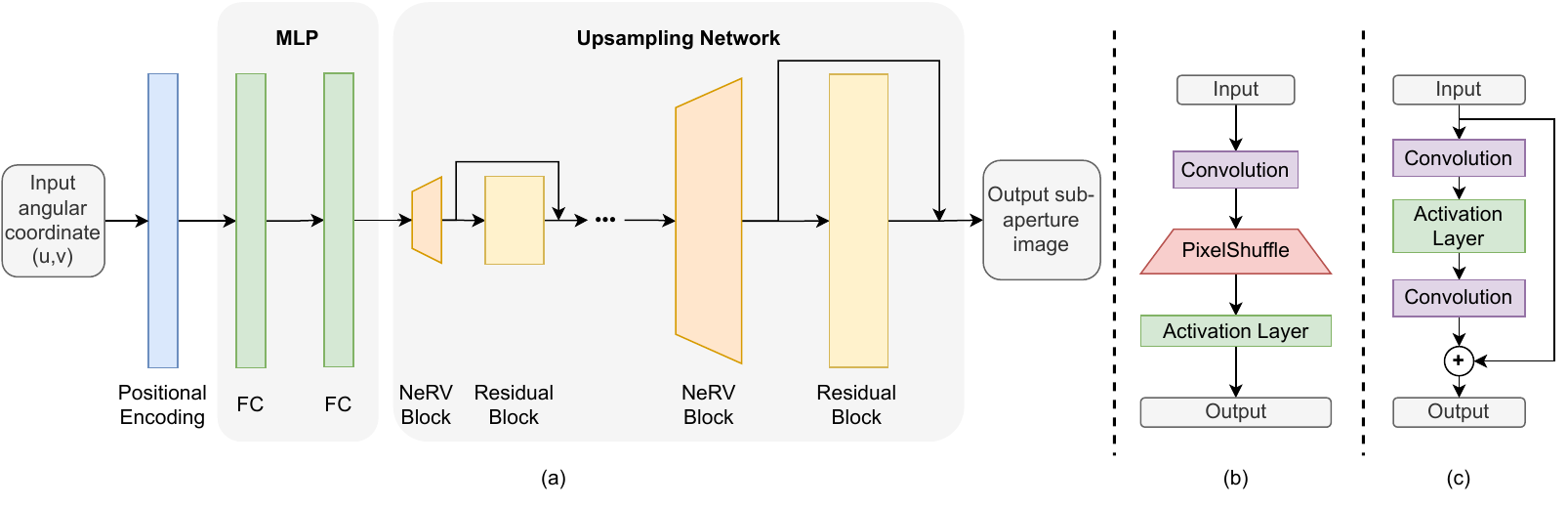}
    \caption{(a) Model architecture of our proposed method. (b) NeRV block structure\cite{chen2021nerv}. (c) Residual block structure\cite{he2016deep}. Our model comprises positional encoding, multi-layer perceptron (MLP) and upsampling network. MLP includes two fully connected (FC) layers. The upsampling network contains 5 consecutive NeRV blocks and residual blocks. Each NeRV block can upscale the height and width of the input feature map by a preset factor using the PixelShuffle technique\cite{shi2016real}. Residual blocks are able to improve output image quality by the residual structure shown in (c)}
    \label{fig:arch}
\end{figure*}

\subsection{Implicit Representation for Light Fields}
\label{sec:INR-LF}
As previously mentioned, a light field image can be regarded as a pseudo video sequence, with each SAI representing a frame. For example, a light field image consisting of 11×11 SAIs can be treated as a pseudo video sequence of 121 frames. NeRV (Neural Representation for Videos)\cite{chen2021nerv} is one of the INR architectures for videos. Given frame index $t$, NeRV network outputs the corresponding RGB image $v_t\in\mathbb{R}^{H\times W\times 3}$. 

Based on NeRV, we design an implicit neural representation network for light fields. The network architecture is shown in Figure \ref{fig:arch}. The input of our proposed network is angular coordinate $(u,v)$ and the output is the corresponding SAI image $L_{(u,v)}\in\mathbb{R}^{H\times W\times 3}$. Compared to pixel-wise implicit representation, SAI-wise representation is more efficient, especially for light field images with large number of pixels.

\subsubsection{Positional Encoding}
As a prevalent technique in INR network design, positional encoding maps the input coordinate to a vector with higher dimensions, which facilitates the network to acquire high-frequency information\cite{mildenhall2020nerf}. In our method, the input angular coordinate $(u,v)$ is normalized to $[-0.1,0.1]$, and then transformed into a high-dimensional vector $\mathcal{E}$ consisting of the Fourier basis
\begin{equation}
    \begin{split}
        \mathcal{E}(u,v)=[
        \sin{(2\pi b^0 u)},\sin{(2\pi b^1 u)},\dots,\sin{(2\pi b^{L-1} u)}, \\ \cos{(2\pi b^0 u)},\cos{(2\pi b^1 u)},\dots,\cos{(2\pi b^{L-1} u)}, \\
        \sin{(2\pi b^0 v)},\sin{(2\pi b^1 v)},\dots,\sin{(2\pi b^{L-1} v)}, \\
        \cos{(2\pi b^0 v)},\cos{(2\pi b^1 v)},\dots,\cos{(2\pi b^{L-1} v)}] \\
        (b\ge1,\ L\in\mathbb{N}),
    \end{split}
\end{equation}
where $b$ and $L$ are hyper-parameters.

\subsubsection{Network Architecture}
As Figure \ref{fig:arch} shows, the input angular coordinate undergoes positional encoding, the MLP network and the upsampling network to generate the corresponding SAI. Directly outputting the final SAI using MLP alone will result in huge parameters; thus, MLP merely generates a preliminary feature map, and then an upsampling network is used to gradually expand its size.

There are multiple NeRV blocks\cite{chen2021nerv} and residual blocks\cite{he2016deep} in the upsampling network. Each NeRV block consists of a convolution layer, a PixelShuffle\cite{shi2016real} layer and an activation layer. The height and width of the feature map can be scaled by a predefined factor after each NeRV block. A residual block is incorporated following each NeRV block to further improve the quality of the output image.

\subsubsection{Loss Objective}
We select L1 and SSIM fusion loss as our loss objective for a balance between objective and subjective performance:
\begin{equation}
    \begin{split}
        Loss=\frac{1}{UV}\sum_{u=0}^{U-1}\sum_{v=0}^{V-1}\alpha||L_{(u,v)}-L'_{(u,v)}||_1 \\ 
        +(1-\alpha)(1-{\rm SSIM}(L_{(u,v)},L'_{(u,v)})),
    \end{split}
\end{equation}
where $U$, $V$ are the horizontal and vertical angular resolution of the light field image, $L_{(u,v)}$ is the ground truth SAI at $(u,v)$, $L'_{(u,v)}$ is the predicted SAI at $(u,v)$, $\alpha$ is a hyper-parameter used to balance these two parts.

\subsection{Model Compression}
\label{sec:modelcomp}
Model compression is implemented to further lower the bitrate while retaining the reconstruction quality once the network is well optimized to overfit the target light field image. The model compression pipeline includes pruning, quantization, and entropy coding.

\subsubsection{Pruning}
Pruning is an efficient strategy for model compression that can increase model sparsity by removing unnecessary weights. In this case, we employ the pruning method proposed by Jaeho Lee et al \cite{lee2021layeradaptive}.It first determines the layer-adaptive magnitude-based pruning (LAMP) score of each weight, which is a rescaled version of the weight magnitude, followed by a global pruning based on the LAMP score
\begin{equation}
	{\hat{w}}_i=\left\{
		\begin{aligned}
			w_{i} & , & \quad{\rm if}\ {\rm score}(w_{i})\geq {\rm score}_{\rm thres} \\
			0 & , & \quad{\rm otherwise},
		\end{aligned}
	\right.
\end{equation}
where $w_{i}$ is the $i$-th original model parameter, ${\hat{w}}_i$ is the $i$-th pruned model parameter, ${\rm score}(w_{i})$ is the LAMP score of the $i$-th original model parameter, and ${\rm score}_{\rm thres}$ is the LAMP score threshold value for all model parameters. A fine-tuning procedure is followed by pruning to help the network regain performance.

\subsubsection{Quantization}
After model pruning, a quantization procedure is carried out to further minimize the number of bits required for network parameters. A network parameter $\delta_i$ (usually FP-32) is mapped to a $b$-bit integer
\begin{equation}
	Q(\delta_i) = {\rm round}(\frac{\delta_{i}-\delta_{\min}}{S}), \quad S=\frac{\delta_{\max}-\delta_{\min}}{2^b},
\end{equation}
where 'round' is rounding the value to the closest integer, $S$ the scale factor, $\delta_{\max}$ and $\delta_{\min}$ the max and min value for the parameter tensor $\delta$. When decoding, parameter $\delta_i$ can be recovered by
\begin{equation}
	\hat{\delta_{i}}=Q(\delta_i)*S+\delta_{\min}.
\end{equation}
	
\subsubsection{Entropy Coding}
Following quantization, adaptive arithmetic coding is adopted as the entropy coding model to generate the final bitstream. Different from the processes above, this one is lossless and can recover the quantized parameters without any reconstruction loss.

\section{Experiments}
In this section, we conduct a series of experiments to verify the effectiveness of our model. First we introduce our experimental setup, including datasets and implementation details. Then experimental results comparing the rate-distortion performance, perceptual quality and decoding time with other mainstream light field compression methods are presented.

\subsection{Setup}
\subsubsection{Dataset}
We test our algorithms on the EPFL Light Field dataset\cite{rerabek2015iso}. Four light field images (I01:Bikes, I02:Danger de Mort, I04:Stone Pillars Outside, I09:Fountain\&Vincent 2) from the dataset are selected for compression. Each of them has a spatial resolution of 625×434 and angular resolution of 15×15. Due to the vignetting effect of the lenslet structure, only central 11×11 SAIs are selected for further processing.

\subsubsection{Implementation details}
To test model performance at different bitrates, models with 5 different sizes are designed by changing the positional encoding dimensions, MLP dimensions and number of CNN channels. The input feature map of the first upsampling layer in each model has a width of 8 and a height of 6. The upsampling factors are 2, 2, 5, 2, 2. Therefore, the network outputs an image of size (640, 480). With cropping at margin pixels, we obtain the final output SAI of size (625, 434).

To best overfit the target light field image, we train the network with the Adam optimizer with learning rate of 5e-4 and a cosine annealing learning rate schedule. SiLU (Sigmoid Linear Unit) is used as the activation function. The batchsize is 1 and training epoch is 1000. The balance factor $\alpha$ in loss function is 0.7. To best reduce the bitrate while preserving reconstruction quality, the prune ratio is set to 0.8 with 200 fine-tuning epochs and the number of bits for quantization is 8.

\subsection{Rate-Distortion Performance Comparison}

\begin{figure*}[htbp]
    \centering
    \subfigure[I01: Proposed-0.017bpp, SIREN-0.022bpp, HEVC Serpentine-0.015bpp]{\includegraphics[width=0.9\textwidth]{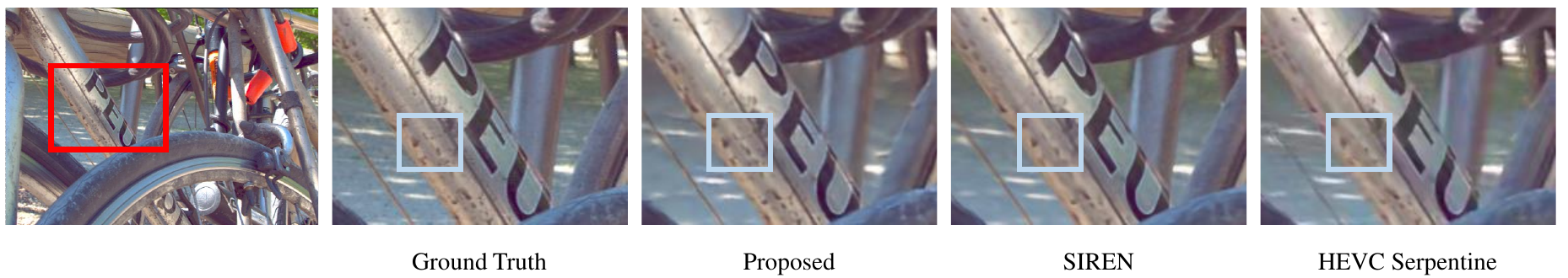}} \\
	\subfigure[I04: Proposed-0.014bpp, SIREN-0.022bpp, HEVC Serpentine-0.014bpp]{\includegraphics[width=0.9\textwidth]{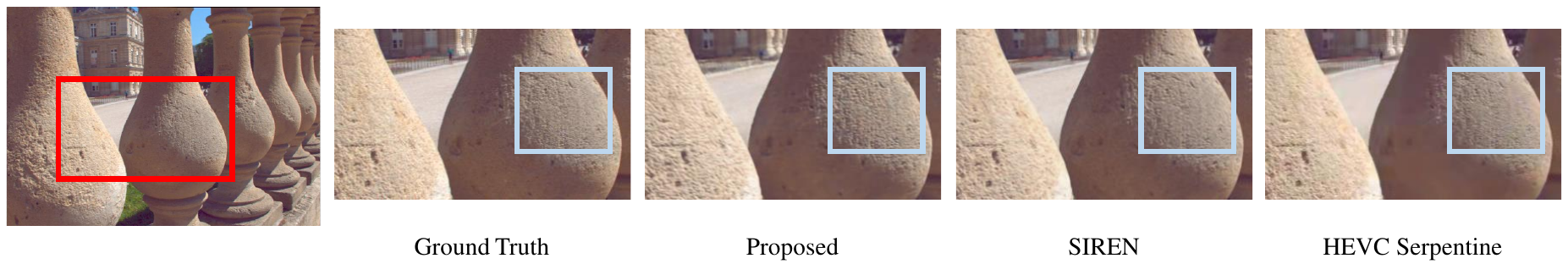}} \\
    \caption{Perceptual quality comparison on (a) I01 and (b) I04. We can observe that our method preserves more details and textures.}
    \label{fig:perceptual}
\end{figure*}

\begin{figure}[h]
    \centering
    \includegraphics[width=0.4\textwidth]{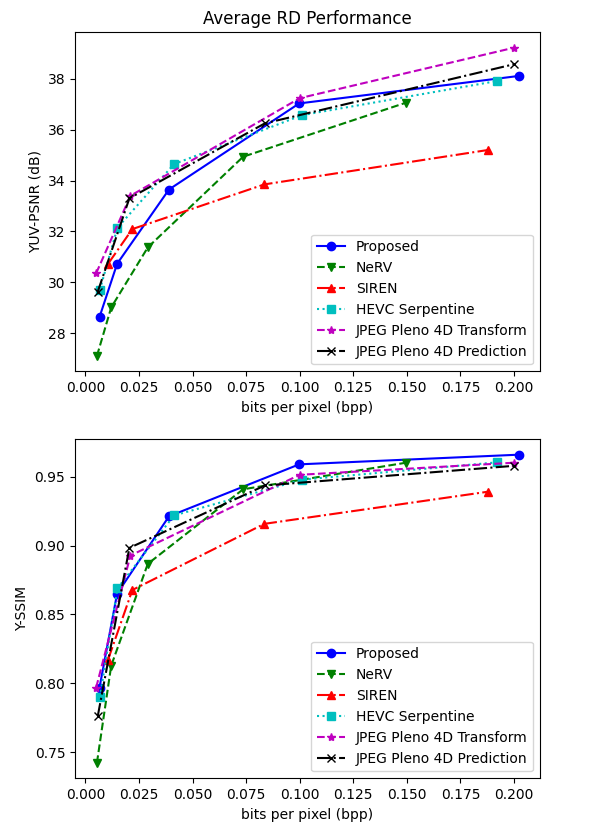}
    \caption{Average RD performance of the proposed method and other methods on I01, I02, I04, I09 of the EPFL LF Dataset.}
    \label{fig:rdcurve}
\end{figure}

\subsubsection{Quality Metrics}
The quality metrics in our experiments are average YUV-PSNR and average Y-SSIM according to the JPEG Pleno document\cite{pereira2019jpeg}. First we convert all reconstructed SAIs from RGB format into YUV444 format, and then YUV-PSNR and Y-SSIM are calculated for each reconstructed SAI. The final PSNRs for the whole light field are computed by averaging the PSNRs of all SAIs, SSIMs are in the same way.

\subsubsection{Other Methods}
We compare the Rate-Distortion (RD) performance of our model with 5 other solutions: the HEVC based pseudo video sequence (PVS) method\cite{vieira2015data}, NeRV\cite{chen2021nerv}, SIREN\cite{sitzmann2020implicit}, JPEG Pleno light field 4D transform coding and prediction coding\cite{schelkens2019jpeg}. 

For the HEVC based PVS method, we arrange all SAIs in serpentine order and encode them with the ffmpeg libx265 medium preset. Constant rate factor (CRF) values are tuned to match the bitrates of our method. The same PVS is also compressed by video neural representation model NeRV. SIREN is a pixel-wise implicit representation model for images. With some slight changes to its input dimensions, SIREN is able to learn the 4-D light field image. Although it was not initially designed for compression, we can also apply the compression pipeline (pruning-quantization-entropy coding) to the SIREN model and test its performance on light field compression. The last two methods are proposed by JPEG Pleno as the standards for light field transform coding and predictive coding. Their results are quoted from the JPEG Pleno document\cite{perra2019performance}.

\subsubsection{Results}
The RD curve of each method is shown in Figure \ref{fig:rdcurve}. From the RD curve, we observe that our method outperforms other methods in terms of Y-SSIM. In terms of YUV-PSNR, our method performs well at high bitrates. However, when bitrates are low, the performance of our method degrades sharply. Two reasons cause this result: first, the super-resolution ability of the upsampling network is greatly restricted by low bitrates; sceond, a relatively small network suffers from a greater performance drop after model compression. The average YUV-PSNR and Y-SSIM of the four tested images and the PSNR \& SSIM drop due to model compression are shown in Table \ref{tab:psnrtable}. Our method also surpasses the original NeRV method which lacks residual connections and only treats the light field as a 1-D pseudo video sequence regardless of its 2-D angular structure.

\begin{table}[htbp]
    \caption{Average Performance and performance drop due to model compression}
    \begin{center}
    \begin{tabular}{|c|c|c|c|c|c|}
    \hline
    bpp & 0.006 & 0.015 & 0.039 & 0.10 & 0.20 \\
    \hline
    YUV-PSNR (dB) & 28.63 & 30.73 & 33.65 & 37.03 & 38.11 \\
    \hline
    PSNR loss (dB) & -0.86 & -0.75 & 0.42 & -0.38 & -0.44 \\
    \hline
    Y-SSIM & 0.797 & 0.865 & 0.921 & 0.959 & 0.966 \\
    \hline
    SSIM loss & -0.027 & -0.018 & -0.006 & -0.002 & -0.002 \\
    \hline
    \end{tabular}
    \label{tab:psnrtable}
    \end{center}
\end{table}

\subsection{Perceptual Comparison}
To assess the perceptual quality of our reconstructed light fields, we extract central SAIs from our method and two other ones for comparison. The result is shown in Figure \ref{fig:perceptual}. It can be observed that at similar bitrates, our method is able to achieve higher perceptual quality. More details and textures can be recovered compared to other methods. However, our method has fading issues. The green background at the top of I01 and the red flowers in front of the building in I04 are faded in our reconstructed images.

\subsection{Decoding Time Comparison}
To show the efficiency of our SAI-wise INR model, we compare our model with pixel-wise model SIREN\cite{sitzmann2020implicit} with decoding time. The experiments are conducted on NVIDIA 2080Ti GPU. Results are shown in Table \ref{tab:timetable}.

\begin{table}[htbp]
    \caption{Average decoding time of the proposed method and SIREN}
    \begin{center}
    \begin{tabular}{|c|c|c|c|c|c|}
    \hline
    bpp (ours) & 0.006 & 0.015 & 0.039 & 0.10 & 0.20 \\
    \hline
    time (s) (ours) & 0.144 & 0.154 & 0.239 & 0.425 & 0.419 \\
    \hline
    bpp (SIREN) & 0.005 & 0.010 & 0.022 & 0.084 & 0.18 \\
    \hline
    time (s) (SIREN) & 2.29 & 2.89 & 4.21 & 7.84 & 10.12 \\
    \hline
    \end{tabular}
    \label{tab:timetable}
    \end{center}
\end{table}

\section{Conclusions}
In this paper, we design a SAI-wise implicit neural representation model for light fields and then use INR formats to compress the light fields. Our model uses the SAI coordinate as the input and outputs the specific SAI of that coordinate. The decoder may utilize any angular coordinate as a query to generate the corresponding SAI instead of reconstructing the whole light field, which guarantees its region of interest (ROI) functionalities. With the help of implicit representation, our method is able to acquire both efficiency and flexibility. In the future, we will incorporate view synthesis into our compression pipeline and investigate new model compression methods that are more suitable for INR network compression tasks.

\bibliographystyle{IEEEtran}
\bibliography{references.bib}

\end{document}